\documentstyle[aps]{revtex}

\begin{document}
\draft
\title{Interference of Resonance Fluorescence from two 
four-level atoms}
\author{T. Wong, S.M. Tan, M.J. Collett and D.F. Walls}
\address{Department of Physics,\\
University of Auckland, Private Bag 92019, Auckland, New 
Zealand}
\maketitle

\begin{abstract}
In a recent experiment by Eichmann{\em \ et al.,} 
polarization-sensitive
measurements of the fluorescence from two four-level ions 
driven by a
linearly polarized laser were made. Depending on the 
polarization chosen,
different degrees of interference were observed. We carry 
out a theoretical
and numerical study of this system, showing that the 
results can largely be
understood by treating the atoms as independent radiators 
which are
synchronized by the phase of the incident laser field. The 
interference and
its loss may be described in terms of the difference 
between coherent and
incoherent driving of the various atomic transitions in the 
steady-state. In
the numerical simulations, which are carried out using the 
Monte Carlo wave
function method, we remove the assumption that the atoms 
radiate
independently and consider the photodetection process in 
detail. This allows
us to see the total interference pattern build up from 
individual
photodetections and also to see the effects of 
superfluorescence, which
become important when the atomic separation is comparable 
to an optical
wavelength. The results of the calculations are compared 
with the
experiment. We also carry out simulations in the non 
steady-state regime and
discuss the relationship between the visibility of the 
interference pattern
and which-path considerations.
\end{abstract}


\preprint{Version: Auckland, \today}

\section{Introduction}

Young's two slit experiment is the canonical experiment 
demonstrating the
wave nature of light. With the addition of a measuring 
device to determine
which slit the photon has passed through, the phenomenon of 
wave-particle
duality can be investigated. Many such path detection 
schemes have been
proposed in the past, such as the Einstein recoiling slit 
\cite{Einstein}
and Feynman's light microscope \cite{Feynman}.

Recently, interest has arisen in the interference of the 
fluorescent light
from two driven atoms which play the role of the slits in 
Young's
experiment. Various situations have been theoretically 
studied, using
two-level and four-level atoms, and the possibility of 
placing the atoms in
a cavity has also been considered.\cite
{many_refend,many_refstart,many_refmid}

In a recent experiment performed by Eichmann {\em et al.}
\cite{Eichmann},
the light scattered by two $^{198}{\rm Hg}^{+}$ ions in a 
trap was observed.
In this experiment, a linearly-polarized traveling wave 
coherent field was
used to drive the ions and the scattered light was observed 
in two
orthogonal polarizations. An interference pattern is 
observed in one case
but not in the other. These results were explained in two 
ways, firstly in
terms of a which-path argument based on ways in which a 
single photon can
interact with the two atoms and secondly in terms of a 
theoretical analysis
by Polder{\em \ }and Schuurmans{\em \ }\cite{Polder} who 
considered coherent
driving of a {\em single} four-level atom.

In this paper, we carry out an analysis which considers the 
presence of both
atoms and stress that for the coherent steady-state 
excitation as used in
the experiment, the which-path argument is not applicable. 
In sections \ref
{sec:analy} to \ref{sec:expt}, we extend the analysis of 
Polder{\em \ }and
Schuurmans to the experimental situation. In brief, the 
atomic dipoles are
only coherently driven in a specific direction given by the 
polarization of
the incident light field, and so interference is only 
expected for scattered
light of that polarization. Scattered light of the 
orthogonal polarization
is also present, since the atoms can decay spontaneously, 
but this is
incoherent with the driving field and does not lead to 
interference.

Besides the above analysis which is carried out using the 
master equation, a
Monte Carlo wave function simulation was also carried out, 
as is discussed
in sections~\ref{sec:mcwf} and~\ref{sec:num}. We use the 
method to simulate
the system without resorting to the assumptions of large 
atomic separation
and independent radiators used in the previous analytical 
sections. When a
fluorescent photon is detected as traveling in a given 
direction, it may not
be possible in general to determine from which atom it 
arose, and this leads
to additional ``dipole-dipole'' type of terms which become 
important when
the atomic separation is comparable to the optical 
wavelength. At small
separations, the atoms act cooperatively giving rise to 
superfluorescence
whereas for large separations, we smoothly recover the 
results for
independent radiators.

In order to compare our results with the experiment, we 
include the effects
of the classical motion of the ions within the trap 
(Sec.~\ref
{sec:class_motion}). As expected, this reduces the 
visibility of the
interference patterns, bringing the results more closely to 
those observed.

In the last section (Sec.~\ref{sec:one_photon}) we consider 
a situation
where the two atom system is re-prepared to its initial 
state after each
individual photon has been detected. This allows us to 
study the transient
regime of this system and we find that interference is 
present in both
polarizations, although one of these disappears in the 
steady-state.
Depending on the initial state however, the visibility of 
this transient
pattern varies, and a which-path argument may be used in 
this case to
explain the results.

\section{The Model}

\label{sec:analy}The atoms are modeled as four-level atoms 
(see fig.~\ref
{fig_4Latom}) interacting with a linearly polarized light 
field polarized
along the $z$ axis and traveling along the $y$ direction. 
We treat the
incident laser light as a classical field and the external 
motion of the
atoms will be neglected. The separation between the atoms 
will be considered
to be large (many wavelengths of the laser light) so that 
dipole-dipole
interactions can be ignored. In this approximation the 
atoms can be
considered as independent fixed sources except that their 
radiation is
synchronized by the definite phase of the incident laser 
light. The atomic
dipole operator ${\bf \vec{\mu}}$ is the sum of atomic 
raising ${\bf \mu }%
^{\uparrow }$ and lowering ${\bf \mu }^{\downarrow }$ 
operators whose
components are given by \cite{Polder} 
\begin{eqnarray}
{\bf \mu }_{x}^{\downarrow } &=&\mu (|1\rangle \langle 
4|+|2\rangle \langle
3|){\bf \hat{x}}  \label{dip01} \\
{\bf \mu }_{y}^{\downarrow } &=&-\imath \mu (|1\rangle 
\langle 4|-|2\rangle
\langle 3|){\bf \hat{y}} \\
{\bf \mu }_{z}^{\downarrow } &=&\mu (|2\rangle \langle 
4|-|1\rangle \langle
3|){\bf \hat{z},}  \label{dip03}
\end{eqnarray}
where ${\bf \mu }_{k}^{\downarrow }$ is the {\it k} 
component of the atomic
dipole, $\mu $ is the dipole matrix element, ${\bf \hat{x}}
$, ${\bf \hat{y}}$
and ${\bf \hat{z}}$ are the usual Cartesian unit vectors. 
The atomic
operator $|i\rangle \langle j|$ couples the $i$'th and 
$j$'th levels of the
atom. The system Hamiltonian for one of the atoms in the 
interaction picture
is 
\begin{equation}
{\bf H}_{s}=\frac{1}{2}\hbar \Delta \left( \sigma _{44}
+\sigma _{33}-\sigma
_{22}-\sigma _{11}\right) +\hbar \Omega \left( \sigma _{24}
+\sigma
_{42}-\sigma _{13}-\sigma _{31}\right) ,
\end{equation}
where $\Delta $ is the detuning between the atomic 
transition and the
incident laser light. For simplicity, the ground states 
(and the excited
states) are considered to be degenerate in energy (see 
fig.~\ref{fig_4Latom}%
). We define the Rabi frequency $\Omega =\mu E_{o}/2\hbar $ 
where $E_{o}$ is
the electric field amplitude of the incident laser and we 
use the notation $%
\sigma _{ij}\equiv |i\rangle \langle j|.$ The equation of 
motion for the
reduced density operator $\rho _{s}$ for the single atom is 

\begin{equation}
\frac{d\rho _{s}}{dt}=\frac{\imath }{\hbar }\left[ \rho _{s}
,{\bf H}%
_{s}\right] +{\cal L}_{\text{relax}}(\rho _{s}),
\end{equation}
where ${\cal L}_{\text{relax}}$ is the relaxation 
superoperator

\begin{eqnarray*}
{\cal L}_{\text{relax}} &=&-\gamma \left( \sigma _{44}\rho 
_s+\rho_s\sigma
_{44}+\sigma _{33}\rho _s+\rho _s\sigma _{33}\right) \\
&&+\gamma \left( \sigma _{14}\rho _s\sigma _{41}+\left( 
\sigma_{24}-\sigma
_{13}\right) \rho _s\left( \sigma _{42}-\sigma _{31}\right) 
+\sigma_{23}\rho
_s\sigma _{32}\right) ,
\end{eqnarray*}
and we have assumed that both excited states have the same 
decay rate $%
\gamma .$

\section{The Optical Bloch Equations}

\label{sec:Opt_Bloch}There are sixteen optical Bloch 
equations corresponding
to the sixteen atomic operators. However, due to the 
geometry chosen for the
incident light, the sixteen equations de-couple into two 
sets of eight. The
first set of coupled equations involve the populations and 
the ``linear
coherences'' ($\langle \sigma _{13}\rangle $ and $\langle 
\sigma
_{24}\rangle $ terms) which relate to the $z$ component of 
the atomic dipole
(see eqn.~\ref{dip03}). 
\begin{eqnarray}
\frac{d}{dt}\langle \sigma _{13}\rangle &=&(-\gamma -\imath 
\Delta )\langle
\sigma _{13}\rangle -\imath \Omega (\langle \sigma _{33}
\rangle -\langle
\sigma _{11}\rangle ) \\
\frac{d}{dt}\langle \sigma _{31}\rangle &=&(-\gamma +\imath 
\Delta )\langle
\sigma _{31}\rangle +\imath \Omega (\langle \sigma _{33}
\rangle -\langle
\sigma _{11}\rangle ) \\
\frac{d}{dt}\langle \sigma _{24}\rangle &=&(-\gamma -\imath 
\Delta )\langle
\sigma _{24}\rangle -\imath \Omega (\langle \sigma _{22}
\rangle -\langle
\sigma _{44}\rangle ) \\
\frac{d}{dt}\langle \sigma _{42}\rangle &=&(-\gamma +\imath 
\Delta )\langle
\sigma _{42}\rangle +\imath \Omega (\langle \sigma _{22}
\rangle -\langle
\sigma _{44}\rangle ) \\
\frac{d}{dt}\langle \sigma _{11}\rangle &=&-\imath \Omega 
(\langle \sigma
_{31}\rangle -\langle \sigma _{13}\rangle )+\gamma (\langle 
\sigma
_{33}\rangle +\langle \sigma _{44}\rangle ) \\
\frac{d}{dt}\langle \sigma _{22}\rangle &=&-\imath \Omega 
(\langle \sigma
_{24}\rangle -\langle \sigma _{42}\rangle )+\gamma (\langle 
\sigma
_{33}\rangle +\langle \sigma _{44}\rangle ) \\
\frac{d}{dt}\langle \sigma _{33}\rangle &=&-\imath \Omega 
(\langle \sigma
_{13}\rangle -\langle \sigma _{31}\rangle )-2\gamma \langle 
\sigma
_{33}\rangle \\
\frac{d}{dt}\langle \sigma _{44}\rangle &=&-\imath \Omega 
(\langle \sigma
_{42}\rangle -\langle \sigma _{24}\rangle )-2\gamma \langle 
\sigma
_{44}\rangle .
\end{eqnarray}
Since normalization requires that

\begin{equation}
\sum_{i=1}^4\langle \sigma _{ii}\rangle =1,
\end{equation}
the number of independent equations may be reduced to 
seven. Setting the
time derivatives to zero for the steady-state 
yields\footnote{%
These solutions are compatible to those of ref~\cite{Polder}
 since only one
atom has been included at this time.} 
\begin{eqnarray}
\langle \sigma _{13}\rangle _{ss} &=&(\overline{\Delta }
+\imath \overline{%
\gamma })\langle \sigma _{33}\rangle _{ss} \\
\langle \sigma _{31}\rangle _{ss} &=&\langle \sigma _{13}
\rangle _{ss}^{*} \\
\langle \sigma _{24}\rangle _{ss} &=&(-\overline{\Delta }
-\imath \overline{%
\gamma })\langle \sigma _{44}\rangle _{ss} \\
\langle \sigma _{42}\rangle _{ss} &=&\langle \sigma _{24}
\rangle _{ss}^{*} \\
\langle \sigma _{11}\rangle _{ss} &=&\frac 12-\langle 
\sigma _{33} \rangle
_{ss} \\
\langle \sigma _{22}\rangle _{ss} &=&\langle \sigma _{11}
\rangle _{ss} \\
\langle \sigma _{33}\rangle _{ss} &=&\frac 
12(\overline{\Delta }^2+ 
\overline{\gamma }^2+2)^{-1} \\
\langle \sigma _{44}\rangle _{ss} &=&\langle \sigma _{33}
\rangle _{ss}
\end{eqnarray}
where $\overline{\Delta }=\Delta /\Omega $ and 
$\overline{\gamma }= \gamma
/\Omega $. As expected from the symmetry of the situation, 
the populations
of the two ground states are equal as are the populations 
of the two excited
states.

A graph of the populations versus laser light strength is 
shown in fig.~\ref
{fig_popcoh}(a). For strong fields the atom is saturated 
with all the
populations equal to one quarter. For weak fields, the 
excited states are
close to zero with the ground states close to one half. The 
linear
coherences are shown in fig.~\ref{fig_popcoh} (b). Since 
the coherences are
purely imaginary we plot the imaginary part of $\langle 
\sigma _{13}\rangle $
and $\langle \sigma _{24}\rangle $. We see that as the 
field strength
increases the coherences grow and reach a peak for $\Omega 
/\gamma \approx
2/3$. There is a gradual loss of coherence as the field is 
further increased
due to the saturation of the atom. At low field strengths 
the atom spends
most of its time in the ground state and is very rarely 
pumped to the
excited states. It behaves like a coherently driven linear 
oscillator and
scatters coherently. In the strong field case, the atom 
undergoes Rabi
oscillations and is frequently pumped to the excited states 
from which it
spontaneously decays. These spontaneous emission events 
destroy the
coherences.

The second set of equations consist of the ``cross 
coherences'' ($\langle
\sigma _{14}\rangle $ and $\langle \sigma _{23}\rangle $) 
and the coherences
between the ground $\langle \sigma _{12}\rangle $ and 
between the excited
states $\langle \sigma _{34}\rangle $. 
\begin{eqnarray}
\frac{d}{dt}\langle \sigma _{14}\rangle &=&(-\gamma -\imath 
\Delta )\langle
\sigma _{14}\rangle -\imath \Omega (\langle \sigma _{12}
\rangle +\langle
\sigma _{34}\rangle ) \\
\frac{d}{dt}\langle \sigma _{23}\rangle &=&(-\gamma -\imath 
\Delta )\langle
\sigma _{23}\rangle +\imath \Omega (\langle \sigma _{21}
\rangle +\langle
\sigma _{43}\rangle ) \\
\frac{d}{dt}\langle \sigma _{12}\rangle &=&-\imath \Omega 
(\langle \sigma
_{14}\rangle +\langle \sigma _{32}\rangle )-\gamma \langle 
\sigma
_{34}\rangle \\
\frac{d}{dt}\langle \sigma _{34}\rangle &=&-\imath \Omega 
(\langle \sigma
_{32}\rangle +\langle \sigma _{14}\rangle )-2\gamma \langle 
\sigma
_{34}\rangle
\end{eqnarray}
The other four equations for $\langle \sigma _{41}\rangle $,
 $\langle \sigma
_{32}\rangle $, $\langle \sigma _{21}\rangle $ and $\langle 
\sigma
_{43}\rangle $ are the Hermitian conjugates of these 
equations. Solving this
set of equations in the steady-state we find that only the 
trivial solution
remains so that all these coherences ultimately vanish. Any 
initial
coherences involving these terms are damped in the 
transient regime. Since
the cross coherences, $\langle \sigma _{14}\rangle $ and 
$\langle \sigma
_{23}\rangle $, determine the $x$ and $y$ components of the 
atomic dipole
(see eqn.~\ref{dip01}-\ref{dip03}), there is no 
interference from light
polarized in the $xy$ plane.

\section{The Interference Pattern}

We have calculated in the previous section the steady-state 
solutions for
the atomic coherences and populations for the case of a 
single four-level
atom interacting with a classical laser light field 
linearly polarized along
the $z$ axis. In this section we wish to use these to 
calculate the
far-field interference pattern from two such atoms in the 
case when their
separation is large enough that they may be treated 
independently.

Let us consider a screen placed in the far-field (large 
$y$) and oriented in
the $xz$ plane. In the far-field, and in the paraxial 
approximation, the
electric field at a point on the screen at time $t$ due to 
an oscillating
dipole is proportional to the projection onto the plane of 
the screen of the
acceleration of the dipole moment at the retarded time 
$t-\tau ,$where $\tau$
is the light travel time from the dipole to the observation 
point. We shall
label a point on the screen by $(\tau _1,\tau _2)$ where 
these are the light
travel times from each of the atoms to the point on the 
screen. The
intensity of the light at this point is 
\begin{equation}
I(\tau _1,\tau _2)\propto \left\langle E_x^{\uparrow } 
E_x^{\downarrow
}+E_z^{\uparrow }E_z^{\downarrow }\right\rangle ,
\end{equation}
where 
\begin{equation}
E_k^{\uparrow }(t;\tau _1,\tau _2)\propto \text{{\rm e}}
^{-i\omega (t-\tau
_1)}\mu _k^{\uparrow }+\text{{\rm e}}^{-i\omega (t-\tau _2)}
 {\cal U}%
_k^{\uparrow },
\end{equation}
for $k\in \left\{ x,z\right\} ,$ $\mu $ and ${\cal U}$ are 
the atomic
dipoles of the first and second atoms respectively and 
$\omega $ is the
angular frequency of the laser light. If we use the 
notation $\sigma _{ij}$
for the atomic operators of the first atom and $\Sigma _{ij}
$ for those the
second, we note that the cross terms like $\langle \mu 
_x^{\uparrow }{\cal U}%
_x^{\downarrow }\rangle $ are zero in the steady-state 
since 
\begin{eqnarray}
\langle \mu _x^{\uparrow }{\cal U}_x^{\downarrow }\rangle 
_{ss} &\propto
&\langle (\sigma _{41}+\sigma _{32})(\Sigma _{14}+\Sigma 
_{23}) \rangle _{ss}
\nonumber \\
&=&(\langle \sigma _{41}\rangle _{ss}+\langle \sigma _{32}
\rangle
_{ss})(\langle \Sigma _{14}\rangle _{ss}+\langle \Sigma 
_{23} \rangle _{ss})
\nonumber \\
&=&0.
\end{eqnarray}
We have factorized the products above since we have assumed 
that the atoms
are independent. The intensity of the interference pattern 
when all the
light is detected is given by 
\begin{eqnarray}
\hspace{-3mm}I_{\text{unpol}}(\tau _1,\tau _2) &\propto 
&\left \langle \mu
_x^{\uparrow }\mu _x^{\downarrow }+{\cal U}_x^{\uparrow }
{\cal U}%
_x^{\downarrow }+\mu _z^{\uparrow }\mu _z^{\downarrow } 
+{\cal U}%
_z^{\uparrow }{\cal U}_z^{\downarrow }\right\rangle  
\nonumber \\
&&+\left\langle \mu _z^{\uparrow }{\cal U}_z^{\downarrow }
\right \rangle
\exp (\imath [\omega (\tau _1-\tau _2)-\phi _o])  \nonumber 
\\
&&+\left\langle {\cal U}_z^{\uparrow }\mu _z^{\downarrow }
\right \rangle
\exp(-\imath [\omega (\tau _1-\tau _2)-\phi _o]),
\end{eqnarray}
where the additional phase term $\phi _o$ has been included 
to allow for the
phase of the incident light to differ at the positions of 
the two ions. If
this is normalized by twice the intensity due to single 
atom fluorescence,
we obtain 
\begin{equation}
I_{\text{N,unpol}}(\tau _1,\tau _2)=1+V_{\text{unpol}}\cos 
[\omega
(\tau_1-\tau _2)-\phi _o],
\end{equation}
where 
\begin{equation}
V_{\text{unpol }}= \frac{1}{2}\left( \frac{\overline{\Delta 
}^2 +\overline{%
\gamma }^2}{\overline{\Delta }^2+\overline{\gamma }^2+2}
\right)
\label{Vis_unpol}
\end{equation}
We note that this interference pattern is the sum of an 
incoherent term and
a coherent term and so the visibility is always less than 
one half. If we
introduce a polarizer at angle $\eta $ to the $z$ axis in 
front of the
screen, it is possible to separate out the two components. 
The expression
for the intensity is then 
\begin{eqnarray}
\hspace{-3mm}I_{\text{pol}}(\tau _1,\tau _2;\eta ) &\propto 
&\left \langle
\mu _x^{\uparrow }\mu _x^{\downarrow }+{\cal U}_x^{\uparrow 
} {\cal U}%
_x^{\downarrow }\right\rangle \sin ^2\eta +\left\langle \mu 
_z^{\uparrow
}\mu _z^{\downarrow }+{\cal U}_z^{\uparrow }{\cal U}
_z^{\downarrow}\right%
\rangle \cos ^2\eta  \nonumber \\
&&+\left\langle \mu _z^{\uparrow }{\cal U}_z^{\downarrow }
\right \rangle
\exp (\imath [\omega (\tau _1-\tau _2)-\phi _o])\cos ^2\eta 
 \nonumber \\
&&+\left\langle {\cal U}_z^{\uparrow }\mu _z^{\downarrow }
\right \rangle
\exp (-\imath [\omega (\tau _1-\tau _2)-\phi _o])\cos 
^2\eta
\end{eqnarray}
Normalizing this as before yields 
\begin{equation}
I_{\text{N,pol}}(\tau _1,\tau _2;\eta )=1+V_{\text{pol}}
(\eta ) \cos
[\omega(\tau _1-\tau _2)-\phi _o]  \label{I_norm}
\end{equation}
where the visibility is now given by 
\begin{equation}
V_{\text{pol}}(\eta )=\left( \frac{\overline{\Delta }^2 
+\overline{\gamma }^2%
}{\overline{\Delta }^2+\overline{\gamma }^2+2}\right) \cos 
^2\eta .
\label{eqn_vis}
\end{equation}
Thus if we detect the $x$ polarized light by setting $\eta 
=90^{\circ },$ we
see that the visibility is zero for any field strength and 
detuning. On the
other hand if the polarizer is set along the $z$ axis 
($\eta =0$) then the
visibility is non-zero. This visibility is plotted against 
field strength in
fig.~\ref{fig_vis} for various detunings with the polarizer 
aligned with the 
$z$ axis ($\eta =0$). The maximum theoretical visibility of 
one occurs for
all detunings in the limit of zero field strength. As the 
field strength is
increased the visibility falls towards zero, with the 
detuning determining
the rate of this drop.

\section{Comparison with experimental parameters}

\label{sec:expt}In the experiment of Eichmann {\em et al.} 
\cite{Eichmann}
the incident laser is linearly polarized at an angle of 
$62^{\circ }$
degrees to the $z$ axis in the $yz$ plane. This gives a $y$ 
component to the
atomic dipole which is not considered in the above 
calculation, but this
does not affect the field on a screen in the $xz$ plane. 
The inclination
also introduces a non-zero phase difference $\phi _o$ 
between the two atoms
which has be included in our calculations. For more general 
orientations of
the incident field, we resort to numerical simulations as 
described later 
\footnote{%
A numerical simulation will be introduced later which can 
simulate the
geometry of the incident light in a more flexible manner.}. 
Another
difference between our analysis and the experiment is that 
instead of
rotating a polarizer in the output field, a fixed Brewster 
plate is used and
the polarization of the incident light is changed. Our 
analysis is not valid
for incident light polarized along the $x$ direction, 
however our use of a
rotating polarizer gives some insight into the 
polarization-sensitive
measurement.

Taking into account the geometry of the experiment, we 
expect that the
measured interference pattern should be 
\begin{equation}
I_{\text{N,unpol }}= 1 + \frac{1}{2}
\left(\frac{\overline{\Delta}^2+%
\overline{\gamma }^2}{\overline{\Delta }^2+\overline{\gamma 
}^2+2}\right)
\cos\left[\left(\frac{2\pi d}{\lambda}
\right)\sin(\phi-28^{o})
-\phi_{o}\right]  \label{I_unpol}
\end{equation}
when all the light is collected and 
\begin{equation}
I_{\text{N,pol }}= 1 + \cos ^2\eta\left( 
\frac{\overline{\Delta }^2 +%
\overline{\gamma}^2}{\overline{\Delta }^2+\overline{\gamma }
^2+2}
\right)\cos\left[\left(\frac{2\pi d}{\lambda}\right)
\sin(\phi-28^{o})-\phi_{o}\right]
\end{equation}
when a polarization-sensitive measurement is made. Where 
$d$ is the distance
between the atoms, $\phi$ is angle relative to the incident 
light and $%
\lambda$ is the wavelength of this light.

The visibility measured in the experiment when all the 
light was collected
had a value of $0.2$. The maximum visibility from 
eqn.~\ref{I_unpol} is one
half. This less than ``ideal'' measured visibility is not 
surprising since
many detrimental factors have not been included in the 
analytical treatment.
Factors such as stray light entering the detector and 
quantum jumps to other
levels in the Mercury ions so that we may observe only the 
fluorescence from
one ion.\cite{Itano} The motion of the ions in the trap 
also smear out the
interference pattern, this will be discuss in detail in a 
later section.

\section{Numerical simulation of the system}

\label{sec:mcwf}In this section, we describe the use of a 
Monte Carlo wave
function method for numerical simulation of the system. 
This approach allows
greater flexibility in specifying the geometry of the 
incident light and
also allows us to include the photodetection process as 
part of the
simulation. By doing this, we no longer simply relate the 
far field
intensity of the fluorescent light to the moments of the 
atomic variables,
but explicitly consider how atomic quantum jumps introduce 
photons into
electromagnetic field modes propagating in various 
directions. The advantage
of this approach is that it allows the atoms to be 
separated by arbitrary
distances and takes into account the dipole-dipole 
correlations between the
atoms. This simulation approach is very closely related to 
the actual
experimental situation as it gives a time-resolved 
classical record of
photocounts at different angles which gradually build up 
into an
interference pattern.

For simplicity, we assume idealized unit efficiency 
photodetectors in the
far field covering the entire $4\pi $ steradians 
surrounding the two atoms.
Each detector is assumed to be polarization-sensitive and 
resolves detected
photons into two orthogonal linear polarizations. Working 
in spherical polar
coordinates with polar angle $\theta $ and azimuthal angle 
$\phi$ (see fig.~%
\ref{coord_sys}), we use the unit vectors $\hat{\epsilon}
_\theta $ and $\hat{%
\epsilon}_\phi $ as a basis for the polarization of the 
emitted photon.
These unit vectors are related to Cartesian coordinates by 
\begin{eqnarray}
\hat{\epsilon}_\theta &=&\cos \theta \cos \phi {\bf \hat{x}}
+\cos \theta
\sin \phi {\bf \hat{y}}-\sin \theta {\bf \hat{z}} \\
\hat{\epsilon}_\phi &=&-\sin \phi {\bf \hat{x}}+\cos \phi 
{\bf \hat{y}}
\end{eqnarray}
where ${\bf \hat{x}}$, ${\bf \hat{y}}$ and ${\bf \hat{z}}$ 
are the Cartesian
unit vectors. The system Hamiltonian for the two atoms is 
just the sum of
the individual Hamiltonians for each atom 
\begin{eqnarray}
{\rm H}_{sys} &=&\frac 12\hbar \Delta \left( \sigma _{44}
+\sigma
_{33}-\sigma _{22}-\sigma _{11}+\Sigma _{44}+\Sigma _{33}
-\Sigma
_{22}-\Sigma _{11}\right)  \nonumber \\
&&+\hbar \Omega _z\left( \sigma _{24}+\sigma _{42}-\sigma 
_{13} -\sigma
_{31}+\Sigma _{24}+\Sigma _{42}-\Sigma _{13}-\Sigma _{31}
\right)  \nonumber
\\
&&+\hbar \Omega _{+}\left( \sigma _{41}+\sigma _{23}+\Sigma 
_{41} +\Sigma
_{23}\right) +\hbar \Omega _{-}\left( \sigma _{14}+\sigma 
_{32} +\Sigma
_{14}+\Sigma _{32}\right) ,
\end{eqnarray}
where we have set $\Omega _{\pm }=\Omega _x\pm \imath 
\Omega _y,$ and where $%
\Omega _x$, $\Omega _y$ and $\Omega _z$ are the Rabi 
frequencies
corresponding to the $x,$ $y,$ and $z$ components of the 
electric field 
\begin{equation}
\Omega _k=\frac{\mu \vec{E}_k}{2\hbar },\hspace{15mm}k=x,y,
z,
\end{equation}
where $\mu$ is the dipole matrix element and each atom is 
considered to feel
the same electric field $\vec{E}$. We are assuming here 
that the incident
light is propagating at right angles to the line joining 
the two atoms. The
relaxation superoperator is in Lindblad form 
\begin{equation}
{\cal L}_{relax}(\rho _s)=-\frac 12\sum_m\left( C_m^{\dag }
C_m\rho_s+\rho
_sC_m^{\dag }C_m\right) +\sum_mC_m\rho _sC_m^{\dag },
\end{equation}
where the collapse operators $C_m$ correspond to the 
couplings between
system and the baths. When the atoms are far apart, they 
may be regarded as
coupling to independent baths and so collapses associated 
with one atom may
be distinguished from those associated with the other. If 
we explicitly
include the photodetection scheme, the appropriate collapse 
operators are
parameterized by the direction and polarization of the 
outgoing photon. For
spontaneous emission from a single two-level atom, the 
appropriate collapse
operators (including recoil) are \cite{Molmer} 
\begin{equation}
C_{\Omega ,\hat{\epsilon}}=\left( \frac{3\gamma }{8\pi }
\right) ^{1/2}\exp
\left( -\imath \vec{{\bf k}}\cdot \vec{{\bf R}}\right) 
\left( \hat{\epsilon}%
^{*}\cdot \vec{{\bf S}}^{\downarrow }\right) ,
\end{equation}
where $\vec{{\bf R}}$ is the atomic position operator, 
$\vec{{\bf k}}$ is
the wave number of the outgoing photon traveling in 
direction $\Omega $, $%
\vec{{\bf S}}^{\downarrow }$ is an operator proportional to 
the atomic
lowering operator ($\vec{{\bf S}}^{\downarrow }\propto 
\sigma^{-}$) and $%
\hat{\epsilon}$ labels the outgoing polarization. For our 
situation with two
fixed atoms, $\vec{{\bf R}}$ may be replaced by the 
locations of the two
atoms $\vec{{\bf r}}_1$ and $\vec{{\bf r}}_2$ (see 
fig.~\ref{coord_sys}) and
both atoms couple to each mode so that 
\begin{equation}
C_{\Omega ,{\bf \epsilon }}={\cal N}\left\{ \exp \left( 
-\imath \vec{{\bf k}}%
\cdot \vec{{\bf r}}_1\right) \left( \hat{\epsilon}^{*}\cdot 
\vec{{\bf S}}%
_1^{\downarrow }\right) +\exp \left( -\imath \vec{{\bf k}}
\cdot \vec{{\bf r}}%
_2\right) \left( \hat{\epsilon}^{*}\cdot \vec{{\bf S}}
_2^{\downarrow}\right)
\right\}
\end{equation}
where ${\cal N}$ is a normalization factor. $\vec{{\bf S}}
_1^{\downarrow }$
and $\vec{{\bf S}}_2^{\downarrow }$ are proportional to the 
atomic dipole
operators for the first and second atoms respectively. The 
Cartesian
components of these operators are 
\begin{eqnarray}
\vec{{\bf S}}_1^{\downarrow } &\propto &\sigma _x{\bf 
\hat{x}} +\sigma _y%
{\bf \hat{y}}+\sigma _z{\bf \hat{z}} \\
\vec{{\bf S}}_2^{\downarrow } &\propto &\Sigma _x{\bf 
\hat{x}} +\Sigma _y%
{\bf \hat{y}}+\Sigma _z{\bf \hat{z}}
\end{eqnarray}
where the $\sigma _i$'s are given by 
\begin{eqnarray}
\sigma _x &=&\sigma _{14}+\sigma _{23} \\
\sigma _y &=&-\imath \left( \sigma _{14}-\sigma _{23}
\right) \\
\sigma _z &=&\sigma _{24}-\sigma _{13},
\end{eqnarray}
and the $\Sigma _i$'s related to the $\Sigma _{ij}$'s in a 
similar way. We
see explicitly from these expressions that when a quantum 
jump (i.e., a
photodetection) occurs, this may come from either atom, the 
coefficient
being dependent on the polarization and the appropriate 
phase factor.

Between the times of the quantum jumps, the atomic wave 
function is evolved
using the non-Hermitian effective Hamiltonian given by 
\footnote{%
see M$\phi$lmer {\em et al.} \cite{Molmer}} 
\begin{equation}
{\bf H}_{eff}={\bf H}_{sys}-\frac{\imath \hbar }{2}\sum_{m}
C_{m}^{\dag
}C_{m}.
\end{equation}
On the right-hand side, the summation is responsible for 
the decrease in the
norm of the Monte-Carlo wave function. Its expectation 
value is proportional
to the intensity of the fluorescent light and the 
probability that a quantum
jump occurs. For our photodetection model, this eventually 
evaluates to

\begin{eqnarray}
\int \sum_{\hat{\epsilon}\perp \vec{{\rm k}}}C_{\Omega ,
\hat{\epsilon}%
}^{\dag }C_{\Omega ,\hat{\epsilon}}{\rm d}\Omega 
&=&\frac{8\pi }{3}{\cal N}%
^{2}\left\{ \left( \sigma _{x}^{\dag }\sigma _{x}+\sigma 
_{y}^{\dag }\sigma
_{y}+\sigma _{z}^{\dag }\sigma _{z}+\Sigma _{x}^{\dag }
\Sigma _{x}+\Sigma
_{y}^{\dag }\Sigma _{y}+\Sigma _{z}^{\dag }\Sigma _{z}
\right) \right. 
\nonumber \\
&&+{\cal A}(\alpha )\left( \sigma _{x}^{\dag }\Sigma _{x}
+\sigma _{y}^{\dag
}\Sigma _{y}+\Sigma _{x}^{\dag }\sigma _{x}+\Sigma _{y}
^{\dag }\sigma
_{y}\right)  \nonumber \\
&&\left. +{\cal B}(\alpha )\left( \sigma _{z}^{\dag }\Sigma 
_{z}+\Sigma
_{z}^{\dag }\sigma _{z}\right) \right\}  \label{big_eqn}
\end{eqnarray}
where 
\begin{eqnarray}
{\cal A}(\alpha ) &=&\frac{3}{8}\left( \frac{4}{\alpha }
\sin \alpha -{\cal T}%
(\alpha )\right) \\
{\cal B}(\alpha ) &=&\frac{3}{4}{\cal T}(\alpha )
\end{eqnarray}
where $\alpha =|\vec{{\bf k}}|\cdot |\vec{{\bf r}}_{2}
-\vec{{\bf r}}_{1}|$
is the $2\pi $ times the separation between the two atoms 
in optical
wavelengths. The function ${\cal T}$ is given by\footnote{%
The function ${\cal T}(\alpha )$ has similar structural 
form to the
dipole-dipole term used by T.G. Rudolph{\em \ et al}. 
\cite{Ficek} and J.%
{\em \ }Guo{\em \ et al.} \cite{Guo} to modify the decay 
rate of the atoms.} 
\begin{equation}
{\cal T}(\alpha )=\frac{4}{\alpha ^{2}}\left( \frac{1}
{\alpha }\sin \alpha
-\cos \alpha \right) .
\end{equation}
The graph of the functions ${\cal A}$ and ${\cal B}$ are 
shown in fig.~\ref
{fig_Talpha}. Note that we have used the scaled parameter 
$\overline{\alpha }%
=\alpha /2\pi $ corresponding to units of wavelengths. At 
$\overline{\alpha }
$ equal to zero where the two atoms coincide, both 
functions are equal to
one. As the separation between the atoms is increased both 
display damped
oscillatory behavior decaying to zero in the limit of large 
separations. $%
{\cal B}$ is damped more rapidly and is almost zero for 
separations of more
than three wavelengths, whereas ${\cal A}$ has larger 
amplitude oscillations
which can still be clearly seen even after six wavelengths. 
Notice that $%
{\cal B}$ corresponds to the size of the correlations 
between the $z$
components of the atomic dipoles whereas ${\cal A}$ 
determines the size of
the correlations between the $x$ and $y$ components. In the 
limit of large
separation between the two atoms, only the first term of 
eqn.~\ref{big_eqn}
remains and we have 
\begin{eqnarray}
\left[ \int \sum_{\hat{\epsilon}\perp \vec{{\rm k}}}
C_{\Omega ,\hat{\epsilon}%
}^{\dag }C_{\Omega ,\hat{\epsilon}}{\rm d}\Omega \right] 
_{\alpha
\rightarrow \infty } &=&{\cal N}^{2}\frac{8\pi }{3}\left( 
\sigma _{x}^{\dag
}\sigma _{x}+\sigma _{y}^{\dag }\sigma _{y}+\sigma _{z}
^{\dag }\sigma
_{z}+\Sigma _{x}^{\dag }\Sigma _{x}+\Sigma _{y}^{\dag }
\Sigma _{y}+\Sigma
_{z}^{\dag }\Sigma _{z}\right)  \nonumber \\
&\equiv &{\cal N}^{2}8\pi \left( \sigma _{44}+\sigma _{33}
+\Sigma
_{44}+\Sigma _{33}\right) .
\end{eqnarray}
This may be compared with the form expected for two atoms 
coupling to
independent baths for which 
\[
\sum_{m}C_{m}^{\dagger }C_{m}=2\gamma \left( \sigma _{44}
+\sigma
_{33}+\Sigma _{44}+\Sigma _{33}\right) 
\]
We see that in the limit of large separation, the atoms 
behave
independently, and that the normalization condition is 
${\cal N}=(\gamma
/4\pi )^{1/2}$. In the opposite limit of the two atoms 
coinciding 
\begin{eqnarray}
\left[ \int \sum_{\hat{\epsilon}\perp \vec{{\rm k}}}
C_{\Omega ,\hat{\epsilon}%
}^{\dag }C_{\Omega ,\hat{\epsilon}}{\rm d}\Omega \right] 
_{\alpha
\rightarrow 0} &=&{\cal N}^{2}\frac{8\pi }{3}\left( \sigma 
_{x}^{\dag
}\sigma _{x}+\sigma _{y}^{\dag }\sigma _{y}+\sigma _{z}
^{\dag }\sigma
_{z}+\Sigma _{x}^{\dag }\Sigma _{x}+\Sigma _{y}^{\dag }
\Sigma _{y}+\Sigma
_{z}^{\dag }\Sigma _{z}\right.  \nonumber \\
&&+\left. \sigma _{x}^{\dag }\Sigma _{x}+\Sigma _{x}^{\dag }
\sigma
_{x}+\sigma _{y}^{\dag }\Sigma _{y}+\Sigma _{y}^{\dag }
\sigma _{y}+\sigma
_{z}^{\dag }\Sigma _{z}+\Sigma _{z}^{\dag }\sigma _{z}
\right)  \nonumber \\
&\equiv &{\cal N}^{2}\frac{8\pi }{3}\left( \left| \sigma 
_{x}+\Sigma
_{x}\right| ^{2}+\left| \sigma _{y}+\Sigma _{y}\right| ^{2}
+\left| \sigma
_{z}+\Sigma _{z}\right| ^{2}\right) .
\end{eqnarray}
This gives four times the decay rate of a single atom 
instead of two times
for the case of large separation since the interactions 
between the atoms
grow as the separation decreases until at zero separation 
the size of these
dipole-dipole terms become of the same order as the 
population terms. This
gives rise to superfluorescence where the two atoms behave 
as one. This is
clearly illustrated by the factorization of the above 
expression into the
coherent sum of the two atoms.

In order to carry out the Monte Carlo simulation, a wave 
function is evolved
using the Schr\"{o}dinger equation with the non-Hermitian 
effective
Hamiltonian until the square of the norm of the wave 
function reaches a
threshold drawn from a uniform distribution lying between 
zero and one. At
this time, it is necessary to select one of the possible 
quantum jumps. The
probability density of detecting a fluorescence photon 
traveling in
direction $(\theta ,\phi )$ with polarization $\epsilon _i$ 
is given by

\begin{equation}
{\cal P}_{\epsilon _i}(\theta ,\phi )=\frac{\left\langle 
C_{\Omega,
\epsilon_i}^{\dag }C_{\Omega ,\epsilon _i}\right\rangle }
{\left \langle
\int\sum_{\epsilon \perp \vec{{\rm k}}}C_{\Omega ,\epsilon }
^{\dag
}C_{\Omega,\epsilon }{\rm d}\Omega \right\rangle } 
\hspace{15mm}i=\theta 
\text{ or }\phi .  \label{prob_dist}
\end{equation}
Conventionally, a Monte Carlo simulation requires many runs 
to be carried
out to obtain the desired statistics of the system from an 
ensemble of
quantum trajectories. For the problem being considered, the 
result of one
quantum trajectory gives the photo-emission record for the 
two atom system,
and we can let this single trajectory run as long as it is 
necessary to
obtain the required number of photons.

\section{Numerical Results}

\label{sec:num}The raw output of these simulations is a 
list of the emission
times, directions and polarizations of each emitted photon. 
The direction is
represented by the two angles $\theta $ and $\phi$ , with 
the polarization
aligned either with the unit vector $\epsilon _\theta $ or 
$\epsilon _\phi $.

The wavelength of the incident laser light used in all of 
the following
numerical simulations was $194nm$. In the first simulation 
we have chosen
parameters which gave maximum visibility of the 
interference pattern. The
laser field was weak, the ratio of Rabi frequency to decay 
rate being $0.2$.
The atomic separation was set at $0.5\mu m$ so that five 
intensity peaks
would be observed as $\theta $ is varied through 
$180^{\circ }$. This gives
a sufficient number of points in order to clearly 
distinguish each peak with
the number of photodetections simulated. The incident light 
is propagating
along the $y$ axis.

The intensity pattern as seen on a sphere around the ions 
is shown in fig.~%
\ref{fig_surf} for incident light polarized in the $z$ 
direction. Since the
system is symmetric about the $z$ axis, fig.~\ref{fig_surf}
(a) shows an
interference pattern as $\theta $ (but not $\phi $)$\;$ is 
varied. Fig.~\ref
{fig_surf}(b) is another simulation in which the incident 
light is linearly
polarized along the $x$ axis. Even though the system is no 
longer symmetric
about the $z$ axis, there is little dependence of the 
interference on the
angle $\phi $. We shall henceforth only plot the 
interference patterns as
functions of $\theta ,$ and integrate over the angle $\phi 
.$ This
integration over $\phi$ modulates the intensity 
distribution with a
sinusoidal envelope since the solid angle integrated over 
at the equator ($%
\theta =90^{\circ }$) is greater that at the poles ($\theta 
=0^{\circ }$ or $%
180^{\circ }$).

For the case of incident light polarized in the $z$ 
direction, fig.~\ref
{graph0}(a) shows the intensity distribution for $\epsilon 
_{\theta }$
polarized light as a function of angle $\theta $ from a 
simulation
consisting of one hundred thousand quantum jumps. Out of 
these, about half
were detected with $\epsilon _{\theta }$ polarization. The 
histogram%
\footnote{%
We have plotted the intensity distributions as histograms 
without the
internal lines for clarity.} is computed using a bin width 
of $1^{\circ }.$
An interference pattern is clearly visible. From eqn. 
\ref{eqn_vis}, we
expect a visibility of $25/27$ or about $0.9$ but the 
visibility measured
from the simulation is approximately $0.7$. However 
eqn.~\ref{eqn_vis} is
only valid in the paraxial approximation and only considers 
photons
polarized in the $z$ direction. When this approximation 
breaks down (i.e.,
when $\theta $ is no longer close to $90^{\circ }$), the 
$\epsilon _{\theta
} $ polarization will include contributions from the 
components of the
atomic dipole other than $\sigma _{z}$. These will reduce 
the total
visibility as we do not expect interference from the $x$ 
and $y$ components
of the atomic dipole which are not coherently driven. If we 
now look at the
orthogonal polarization (the $\epsilon _{\phi }$ 
polarization) as shown in
see fig.~\ref{graph0}(b), there is no interference pattern. 
The $\epsilon
_{\phi }$ polarization is always perpendicular to the $z$ 
axis for any angle 
$\theta $ so only the $x$ and $y$ components of the atomic 
dipoles
contribute.

We now consider the case where the incident laser light is 
polarized along
the $x$ axis instead of the $z$ axis. The histogram for the 
intensity of $%
\epsilon _\theta $ polarized light versus $\theta $ is 
shown in fig.~\ref
{graph1}(a). There is no obvious interference pattern now 
since the $\theta $
polarization consists mainly of the $z$ component of the 
atomic dipole for
angles around $\theta =90^{\circ }$. Small peaks can be 
seen at about $40$
and $140$ degrees, which are due to the $x$ component of 
the atomic dipole
which contribute in larger amounts as the $\theta $ 
deviates significantly
from $90$ degrees. Fig.~\ref{graph1}(b) displays 
interference as expected
since the $\phi $ polarization only consists of the $x$ and 
$y$ components
of the atomic dipole. Since the incident light is linearly 
polarized along
the $x$ axis, the $y$ component of the scattered light is 
not coherent and
the visibility should be less than seen in ~\ref{graph0}
(a). From the
central peaks the visibility is about one third, 
approximately half that for
the case of light polarized in the $z$ direction.

\section{Effects of motion}

\label{sec:class_motion}We have so far ignored the effects 
of the motion of
the ions as this has been secondary to our goal of 
demonstrating how the
visibility of the interference pattern varies with 
polarization. However it
is a significant effect in the experiment performed by 
Eichmann {\em et al.}
We shall model the motion of the ions as harmonic 
oscillators for each of
the three main modes, one involving stretching and two 
involving tilting\cite
{Eichmann,Raizen}. The stretching mode involves motion of 
the atoms along
the $z$ axis whereas the two tilting modes involve motion 
in the $xy$ plane.
Thus the stretching mode alters the distance between the 
ions and the period
of the interference pattern while the tilting modes change 
the angle $\theta$
of the photon emission displacing the interference pattern. 
Both of these
effects smear the interference pattern, reducing the 
visibility.

The temperature of the ions in the trap were around the 
milli-Kelvin region
and the frequency of oscillation of the modes were of order 
of a MHz. By
comparing the thermal kinetic energy of an ion $k_{{\rm B}}
T/2$ with the
separation of the trap energy levels, $\hbar \omega ,$ we 
find that $k_{{\rm %
B}}T/(2\hbar \omega )\approx 10^2$ and so the motion may be 
treated
classically. The ions experience micromotion within the 
trap and the recoil
momentum of the photon emissions may be neglected.

In the actual experiment, the values used for the 
separation between the
ions was around $5\mu m$. These values are about ten times 
larger than the
values used in the simulations. Thus we expect to see over 
fifty peaks over
a range of one hundred and eighty degrees. We carried out a 
simulation with
an ion separation of $3.35\mu m$, where one hundred 
thousand jumps was
performed. In fig.~\ref{fig_motion}(a), we show the results 
for incident
light which is linearly polarized along the $z$ axis, with 
no motion of the
ions. The peaks are poorly resolved and an excessive length 
of time would be
required to obtain sharper resolution for such finely 
separated peaks. Note
that in this graph, the detector is taken to be 
polarization insensitive and
we count all the photons. Fig.~\ref{fig_motion}(a) shows 
the correct period
(nine peaks from $75$ to $105$ degrees) with a visibility 
of about $0.35$
which is lower than the predicted value of $0.46$ using 
eqn.~\ref{Vis_unpol}.

The geometry of the incident light in the simulation 
differs from the that
of the experiment. The light is propagating along the $y$ 
axis in the
simulation whereas the experiment has the incident light at 
an angle of $%
62^{o}$ relative to the $z$ axis. However, the experiment 
does not detect
light over the entire $4\pi$ steradians, only a small 
portion is actually
observed. This observation region is a small disk 
perpendicular to the $y$
axis. By summing over the $\phi$ angle in the simulation 
and considering
only $\theta$ angles close to the $y$ axis we have an 
approximation of the
experimentally observed region of interest.

The micromotion is simulated in two ways. We keep track of 
the oscillatory
motion of the stretch mode starting with some arbitrary 
initial value. Since
we are interested in the long time (steady-state) regime 
the actual initial
value we choose is unimportant. The previous numerical 
procedure is re-used
but the atomic separation is altered for every photon 
emission according to
the displacement of the stretch mode. This is valid because 
the
characteristic time scale of the stretch mode 
$\tau_{\text{stretch}}$ ($\sim
10^{-6}s$) is far longer than the atomic half-life 
$\tau_{1/2}$ ($\sim
10^{-8}$). The steady-state regime is satisfied since the 
duration of a
simulation, $\tau_{\text{sim}}$, producing one hundred 
thousand
photodetections is approximately $10^{-3}s$ which is much 
longer than the
time scale of the stretch mode. The effects of the tilting 
motion is to
shift the interference pattern. Thus during the simulation 
whenever we
calculate a photodetection, the direction of the photon 
needs to be shifted
according to the current displacement of the tilting modes. 
The $x$ and $y$
displacements are randomly generated from a harmonic 
oscillator probability
distribution. In hindsight we could have generated $z$ 
displacements from a harmonic oscillator distribution as
well, however in the simulations we have chosen to keep 
track of the stretch
mode starting from some arbitrary value. Both methods are 
equivalent for the steady-state regime. 
The requirements on the time scales involved 
is

\begin{equation}
\tau_{\text{sim}} \gg \tau_{\text{stretch}} \gg \tau_{1/2}.
\end{equation}

As expected, when classical motion is included in the 
simulation, the
visibility of the interference is reduced, see 
fig.~\ref{fig_motion}(b). We
see an interference pattern with the peaks placed in the 
same positions as
fig.~\ref{fig_motion}(b) but with a reduced visibility of 
about $0.28$. The
visibility for individual peaks degrades as one deviates 
from the central
peak at ninety degrees. This maybe explained by the fact 
that we have
modeled the motion along the $z$ axis as a stretching where 
the
center-of-mass remains stationary, thus the central peak is 
in the same
position whatever the period. As one deviates to the sides, 
a small change
in period has a increasingly larger effect on the position 
of the peak. The
parameters used in this simulation for the motion were: 
stretch frequency $%
\omega_{stretch}/2\pi =\sqrt{3}$ MHz, tilt frequencies 
$\omega _{tilt}/2\pi =%
\sqrt{5}/2$ MHz (for both modes), radial confinement was 
$30nm$ and the
amplitude of the stretch mode was $150nm$. Experimental 
results were shown
in the paper by Eichmann {\em et al. }for this set of 
parameters. The
visibility was about $0.2$ which is less than the value 
obtained in the
simulation. The greater visibility in the simulation is not 
unexpected since
the effects of stray light and the fact that $^{198}{\rm Hg}
^{+}$ ions are
not ideal four-level atoms would further reduce the 
visibility.

\section{The Transient Regime}

\label{sec:one_photon}So far we have investigated the 
steady-state regime
which corresponds to the experimental situation. However 
the transient
regime is also interesting because it is then possible to 
consider the
which-path argument which explains the disappearance of an 
interference
pattern as being due to information of which atom has 
undergone an emission
being somehow recorded in the system. This knowledge can be 
obtained for
some initial atomic states but not others.

In this section, we consider that the two atoms are 
initialized to some
known state and then are exposed to the incident light. 
When the first
fluorescence photon is detected, the experiment is stopped 
and the atoms are
re-initialized to the original state. Over many trials, an 
interference
pattern may develop in the far field, and it is this 
pattern which is of
interest.

Let us first consider the case where the initial joint 
atomic state is $%
|\psi \rangle =|1,1\rangle $ and the incident light is 
polarized in the $z$
direction so that the $1\leftrightarrow 3$ transition is 
driven. When one of
the atoms has been excited to the state $3$, it may decay 
either to the
state $1$ or to the state $2.$ If the decay is to state $1,
$ the final state
of the atoms is again $|1,1\rangle $ and it is not possible 
to tell from
which atom the light came from. We thus expect that the 
detection of light
polarized in the $\epsilon _\theta $ direction which is 
sensitive to
radiation from the $3\rightarrow 1$ transition will exhibit 
interference. On
the other hand, if the excited atom decays to state $2,$ 
the final state of
the atoms is either $|2,1\rangle $ or $|1,2\rangle ,$ 
leaving a record of
the atom responsible for the emission. We thus expect light 
detected with
polarization in the $\epsilon _\phi $ direction to not 
exhibit any
interference at all.

The results of a simulation with this initial state are 
shown in fig. \ref
{graph_tran1}(a) for $\epsilon _\theta $ polarization and 
in fig.~\ref
{graph_tran1}(b) for $\epsilon _\phi $ polarization. The 
atomic separation
was $500nm$ and the ratio of Rabi frequency to decay rate 
is $0.2$ as
before. As expected, a very distinct interference pattern 
can be seen in
(a)\ consisting of five peaks. The visibility is about 
$0.88$, this is quite
close to the expected theoretical value of about $0.9$. For 
the same set of
parameters this transient simulation gives higher 
visibility than the
corresponding steady-state simulation. Also as expected, no 
interference is
seen in (b).

Let us now consider the initial joint atomic state $|\psi 
\rangle
=\left(|1,1\rangle +|1,2\rangle +|2,1\rangle +|2,2\rangle 
\right) /2.$ In
this case, it is not possible to distinguish the atom 
responsible for
emitting the photon whether it is detected in the $\epsilon 
_\theta $ or the 
$\epsilon _\phi $ polarizations. We thus expect to see 
interference for both
polarizations. The result of the simulation for the 
$\epsilon _\phi$
polarization is shown in fig.~\ref{graph_tran2}, confirming 
that this
interference does occur in the transient regime. As 
discussed previously
however, since the driving field is polarized in the $z$ 
direction, this
coherence must approach zero in the steady-state and there 
can be no
steady-state interference pattern, although interference 
does occur in the
transient regime.

We have used extreme parameters so far, for the case of the 
$|1,1\rangle $
initial state we know for certain that each atom is 
initially in a
particular ground state and for the other initial state 
($|\psi \rangle
=\left( |1,1\rangle +|1,2\rangle +|2,1\rangle +|2,2\rangle 
\right) /2$) we
have no initial preference for which ground state each of 
the atoms are in.
In fig. \ref{graph_tran3} results for the $\epsilon _\phi $ 
polarization are
shown for an initial state $|\psi \rangle =\left( 2|1,
1\rangle
+|1,2\rangle+|2,1\rangle +|2,2\rangle \right) /\sqrt{{7}}.$ 
In this
intermediate case, there is some visibility in the 
transient interference
pattern, but this is only about a third, lower than that 
seen in the
previous case.

In the transient regime we expect interference in the 
undriven polarization
which does not appear in the steady-state regime for the 
majority of the
possible initial states. The interference we see here is 
due to the presence
of initial coherences. It is similar to the case of two 
two-level atoms
prepared in the excited states, one would eventually decay 
emitting a photon
in the process. The atoms are then re-prepared in the 
excited states and the
process is repeated to build up an intensity pattern. It is 
not important in
this regime whether there is coherent driving or not but 
only that the atoms
are excited to higher levels thus giving them an 
opportunity to undergo
spontaneous emission.

\section{Summary}

We have analyzed the experiment by Eichmann {\em et al.} 
and presented
theoretical models for understanding the degree of 
interference observed for
different incident and detected polarizations. For a simple 
geometry of the
incident field, it is possible to solve the optical Bloch 
equations
analytically for a single four-level atom in the 
steady-state, and to extend
this result to the case of two atoms, assuming that they 
radiate
independently. From this analysis, the fluorescent light is 
seen to consist
of a coherent component which exhibits interference and an 
incoherent
component which does not. These components may be separated 
out by
polarizers before the light is detected.

In order to model the actual geometry of the source 
polarization used in the
experiment, a numerical technique based on the Monte Carlo 
wave function
method was used. In this method, the photodetection process 
is simulated so
that the interference pattern builds up gradually from a 
series of
detections at various angles. This qualitatively reproduces 
the experimental
results and also allowed us to consider dipole-dipole 
interactions which
become important when the atomic separation is reduced.

To more closely simulate the experimental configuration, 
classical motion of
the ions was included in the simulation. This reduces the 
visibility of the
interference to a value closer to the experimental result. 
The residual
discrepancy is probably due to the approximation of 
modeling the ions as
four-level atoms and experimental factors such as the 
presence of stray
light on the observation screen.

Finally, we demonstrated the formation of a transient 
interference pattern
in situations where there is no steady-state pattern by 
resetting the atomic
state to a specified initial condition after each 
photodetection. In this
case, a which-path argument can be used to understand the 
visibility of the
transient patterns.

\section{Acknowledgments}

We would like to thank W. Itano and D. Wineland for useful 
discussions. This
research was supported by the University of Auckland 
Research Committee and
the Marsden Fund of the Royal Society of New Zealand.

\begin{figure}[tbp]
\caption{Schematic diagram of the four-level atom with 
decay rate $\gamma$
and detuning $\Delta$ from incident laser light of Rabi 
frequency $\Omega$.}
\label{fig_4Latom}
\end{figure}

\begin{figure}[tbp]
\caption{(a) The atomic populations versus field strength. 
The $%
\langle\sigma_{11}\rangle$ and $\langle\sigma_{22}\rangle$ 
curves are
graphed as the solid line (they coincide) with the $\langle
\sigma_{33}\rangle $ and $\langle\sigma_{44}\rangle$ curves 
graphed as the
dashed line (again they coincide). (b) The atomic linear 
coherences versus
field strength. The $\langle\sigma_{13}\rangle$ and 
$\langle\sigma_{42}
\rangle$ curves are graphed as the solid line (they 
coincide) with the $%
\langle\sigma_{31}\rangle $ and $\langle\sigma_{24}\rangle$ 
curves graphed
as the dashed line (again they coincide). Detuning is set 
at zero for both
graphs.}
\label{fig_popcoh}
\end{figure}

\begin{figure}[tbp]
\caption{ Visibility plotted against field strength for 
three detunings. We
have the axis of the polarizer aligned with the $z$ axis so 
that the angle $%
\eta$ is equal to zero. The solid line (a) is the case of 
zero detuning, the
dashed line (b) is $\Delta = 5\gamma$ and the dashed-dotted 
line (c) is for $%
\Delta =10\gamma$.}
\label{fig_vis}
\end{figure}

\begin{figure}[tbp]
\caption{The spherical coordinate system used in the Monte 
Carlo simulations
with the $\epsilon_{\theta}$ and $\epsilon_{\phi}$ unit 
vectors defined. We
also have atoms one and two at positions $\vec{r}_{1}$ and 
$\vec{r}_{2}$
respectively.}
\label{coord_sys}
\end{figure}

\begin{figure}[tbp]
\caption{The function ${\cal A}$ is shown by the solid 
line. The dashed line
is the plot of the function ${\cal B}$ and the parameter 
$\overline{\alpha}$
is the number of wavelengths (of the incident light) 
separating the two
atoms.}
\label{fig_Talpha}
\end{figure}

\begin{figure}[tbp]
\caption{Plots of intensity as a function of both angles 
$\theta$ and $\phi$%
. The incident light is linearly polarized along the $z$ 
axis for (a) and
the $x$ axis for (b). These plots represents the light 
pattern distributed
over the entire sphere around the two atoms.}
\label{fig_surf}
\end{figure}

\begin{figure}[tbp]
\caption{Histogram of the number of detected photons with 
(a) $\theta$ and
(b) $\phi$ polarizations as a function of the angle 
$\theta$. The incident
laser light was linearly polarized along the $z$ axis. The 
detuning was set
to zero, the atomic separation was $0.5$ $\mu m$ and the 
decay rate $%
\gamma=5\Omega$. The number of photons detected was $50,
086$ and $49,914$
for the $\theta$ and $\phi$ polarizations respectively.}
\label{graph0}
\end{figure}

\begin{figure}[tbp]
\caption{Histogram of the number of detected photons with 
(a) $\theta$ and
(b) $\phi$ polarizations as a function of the angle 
$\theta$. The incident
laser light was linearly polarized along the $x$ axis. The 
detuning was set
to zero, the atomic separation was $0.5$ $\mu m$ and the 
decay rate $%
\gamma=5\Omega$. The number of photons detected was $49,
770$ and $50,230$
for the $\theta$ and $\phi$ polarizations respectively.}
\label{graph1}
\end{figure}

\begin{figure}[tbp]
\caption{Histogram of the number of detected photons 
irrespective of there
polarizations as a function of $\theta$ (a). Classical 
motion is included
for (b). Both simulations used atomic separations of $3.35$ 
$\mu m$ and
decay rates of $\gamma=5\Omega$.}
\label{fig_motion}
\end{figure}

\begin{figure}[tbp]
\caption{(a) The $\theta$ and (b) $\phi$ polarized light 
distribution versus
the angle $\theta$. We have calculated $100,000$ quantum 
trajectories in
this transient simulation using the initial state 
$|\psi\rangle=\left(|1,1%
\rangle+|1,2\rangle+|2,1\rangle+|2,2\rangle \right)/2$. 
There were $49,988$
and $50,012$ $\theta$ and $\phi$ polarized photons detected 
respectively.}
\label{graph_tran1}
\end{figure}
\begin{figure}[tbp]
\caption{The $\phi$ polarized light distribution versus the 
angle $\theta$
from the transient simulation using the initial state 
$|\psi\rangle=|1,1%
\rangle$. There were $50,335$ $\phi$ polarized photons from 
the $100,000$
quantum trajectories.}
\label{graph_tran2}
\end{figure}

\begin{figure}[tbp]
\caption{The $\phi$ polarized light distribution versus the 
angle $\theta$
from the transient simulation using the initial state 
$|\psi\rangle=%
\left(2|1,1\rangle+|1,2\rangle+|2,1\rangle+|2,2\rangle 
\right) / \surd{7} $.
There were $50,175$ $\phi$ polarized photons from the $100,
000$ quantum
trajectories.}
\label{graph_tran3}
\end{figure}


\begin{references}
\bibitem{Einstein}  {M. Jammer, {\em The Philosophy of 
Quantum Mechanics}
(Wiley, New York, 1974), p. 121.}

\bibitem{Feynman}  {R. Feynman, R. Leighton, and M. Sands, 
{\em The Feynman
Lectures on Physics} (Addison-Wesley, Reading, MA, 1965), 
Vol. 3.}

\bibitem{many_refend}  {Th. Richter, Opt. Commum. {\bf 80}, 
285 (1991)}

\bibitem{many_refstart}  {P. Kochan, H.J. Carmichael, P.R. 
Morrow and M.G.
Raizen, Phys. Rev. Lett. {\bf 75}, 45 (1995)}

\bibitem{many_refmid}  {R.G. Brewer, Phys. Rev. A, {\bf 52},
 2965 (1995)}

\bibitem{Eichmann}  {U. Eichmann, J.C. Bergquist, J.J. 
Bollinger, J.M.
Gilligan, W.M. Itano, D.J. Wineland and M.G. Raizen, Phys. 
Rev. Lett. {\bf 70%
}, 2359 (1993).}

\bibitem{Polder}  {D. Polder and M.F.H. Schuurmans, Phys. 
Rev. A {\bf 14},
1468 (1976).}

\bibitem{Itano}  {W.M. Itano, J.C. Bergquist, R.G. Hulet 
and D.J. Wineland,
Phys. Rev. Lett. {\bf 59}, 2732 (1987).}

\bibitem{Molmer}  {K. M$\phi $lmer, Y. Castin and J. 
Dalibard, JOSA B {\bf 10%
}, 524 (1993).}

\bibitem{Ficek}  {T.G. Rudolph, Z. Ficek and B.J. Dalton, 
Phys. Rev. A {\bf %
52}, 636 (1995).}

\bibitem{Guo}  {J. Guo and J. Cooper, Phys. Rev. A {\bf 51},
 3128 (1995)}

\bibitem{Raizen}  {M.G. Raizen, J.M. Gilligan, J.C. 
Bergquist, W.M. Itano
and D.J. Wineland, Phys. Rev. A {\bf 45}, 6493 (1992)}
\end{references}
\end{document}